\documentclass[doublecol]{epl2} 
\usepackage[colorlinks=true,linkcolor=blue,urlcolor=blue,citecolor=blue]{hyperref}
\usepackage{amsmath}

\usepackage{mathtools}
\usepackage{mathrsfs}

\usepackage{braket}
\usepackage{amssymb}

\newcommand{\beq}{\begin{equation}}
\newcommand{\eeq}{\end{equation}}
\newcommand{\bea}{\begin{eqnarray}}
\newcommand{\eea}{\end{eqnarray}}

\newcommand{\ofqw}{(q,\omega)}

\newcommand{\rv}{{\bf r}}

\newcommand{\qv}{{\bf q}}

\title{Nonlinear Density Response and Higher Order Correlation Functions in Warm Dense Matter}
\shorttitle{Nonlinear Density Response and Higher Order Correlation Functions} 

\author{Tobias Dornheim\inst{1,2,*} \and Jan Vorberger\inst{2} \and Zhandos A. Moldabekov\inst{1,2}}
\shortauthor{T. Dornheim \etal}

\institute{                    
  \inst{1} Center for Advanced Systems Understanding (CASUS), D-02826 G\"orlitz, Germany\\
  \inst{2} Helmholtz-Zentrum Dresden-Rossendorf (HZDR), D-01328 Dresden, Germany\\
  \inst{*} Electronic mail: t.dornheim@hzdr.de
}

\abstract{
In a recent letter [\textit{Phys.~Rev.~Lett.}~\textbf{125}, 085001 (2020)], Dornheim \textit{et al.}~have presented the first \textit{ab initio} path integral Monte Carlo (PIMC) results for the nonlinear electronic density response at warm dense matter (WDM) conditions. In the present work, we extend these considerations by exploring the relation between the nonlinear response and three-/four-body correlation functions from many-body theory. In particular, this connection directly implies a comparably increased sensitivity of the nonlinear response to electronic exchange--correlation (XC) effects, which is indeed confirmed by our analysis over the entire relevant range of densities ($r_s=0.5,\dots,10$) and temperatures ($\theta=0.01,\dots,4$). Finally, our work suggests the possibility of deliberately probing the nonlinear regime to experimentally probe three- and potentially even four-body correlation functions in WDM.
}

\begin{document}

\maketitle

\section{Introduction}

Warm dense matter (WDM)~\cite{fortov_review}, an extreme state with high temperatures $T\sim10^4-10^8$k and densities comparable or even exceeding the density of solids, constitutes one of the most active frontiers in plasma physics, material science, and related disciplines. In nature, these conditions occur in astrophysical objects such as giant planet interiors~\cite{Militzer_2008,militzer1}, the core of brown dwarfs~\cite{saumon1,becker}, and neutron star crusts~\cite{Chamel2008}. In addition, WDM is highly relevant for cutting-edge technological applications like hot-electron chemistry~\cite{Brongersma2015}, the discovery of novel materials~\cite{Kraus2016,Kraus2017}, and inertial confinement fusion~\cite{hu_ICF}--the latter offering a potential abundance of clean energy upon being fully mastered. For these reasons, WDM is created experimentally in large research facilities around the globe, such as NIF and LCLS in the United States, or the brand-new European X-FEL in Germany. A topical review of experimental techniques has been presented by Falk~\cite{falk_wdm}. 

At the same time, the accurate theoretical description of WDM is most challenging~\cite{wdm_book,new_POP,review} due to the intricate and nontrivial interplay of a) Coulomb correlations (often ruling out perturbation expansions like Green functions~\cite{kwong_prl-00,Kas_PRL_2017}), b) quantum degeneracy effects like Pauli blocking (which prevents semi-classical methods like standard molecular dynamics), and c) thermal excitations (sparking the need of a full thermodynamic description beyond the ground state). These conditions are typically specified by two characteristic parameters that are both of the order of unity~\cite{Ott2018}: the density parameter, also known as the Wigner-Seitz radius, $r_s=\overline{r}/a_\textnormal{B}$ (with $\overline{r}$ and $a_\textnormal{B}$ being the average inter-particle distance and Bohr radius) and the degeneracy temperature $\theta=k_\textnormal{B}T/E_\textnormal{F}$ (with $E_\textnormal{F}=q_\textnormal{F}^2/2$ being the usual Fermi energy~\cite{quantum_theory}, and $q_\textnormal{F}$ the corresponding wave number).

While no single method that is capable to accurately describe all aspects of WDM currently exists, density functional theory (DFT) has emerged as the de-facto work horse of WDM theory as it combines a manageable computational effort with an often tolerable level of systematic error~\cite{moldabekov2021quantum}. At the same time, the accuracy of DFT crucially depends on the employed exchange--correlation (XC) functional~\cite{Clay_PRB_2014,Clay_PRB_2016}, which is an external input and, thus, cannot be obtained within DFT itself. Furthermore, the performance of different XC-functionals at WDM conditions is much less understood compared to the electronic ground state, which makes both the benchmarking of existing functionals~\cite{karasiev_importance,kushal} and the development of new ones~\cite{groth_prl,ksdt,Karasiev_PRL_2018} extremely important.

A second limitation of contemporary WDM theory is that it is often based on linear response theory (LRT), i.e., the assumption of a linear answer of the electrons to an external perturbation~\cite{nolting}. This is certainly understandable, as a thorough theoretical description of nonlinear effects is substantially more complicated than LRT~\cite{rommel1998,Kalman,Kalman_1987,PhysRevB.56.15654,Bergara1999,bergara1997,Rostami_2017}. First results for the nonlinear density response of electrons in the WDM regime have only recently been presented by Dornheim and co-workers~\cite{Dornheim_PRL_2020_ESA,dornheim2021density,dornheim2021nonlinear,dornheim2021nonlinear_ITCF} based on extensive \textit{ab initio} path integral Monte Carlo (PIMC) simulations. While being computationally highly expensive due to the notorious fermion sign problem~\cite{dornheim_sign_problem,dornheim2021fermion}, the PIMC method~\cite{cep,Dornheim_JCP_permutation_cycles,boninsegni1} allows one to obtain exact results within the given Monte Carlo error bars, and thus to obtain an unassailable benchmark for other methods and approximations. In a nutshell, it was shown that nonlinear effects can indeed become important in many situations that are of relevance to contemporary experiments, for example with free-electron lasers~\cite{Fletcher2015,Amann2012}. Moreover, Moldabekov \textit{et al.}~\cite{Moldabekov_CPP_2021, cpp17} have shown that at $r_s\geq 4$ the screened ion potential computed in LRT exhibits a negative minimum which can lead to an effective ion--ion attraction. However, this could be an artifact of LRT, and might vanish when nonlinear effects are properly taken into account. On the other hand, Ashcroft and coworkers have illustrated the possibility of an ion--ion attraction in a non-ideal electronic medium with $r_s>2$ and $\theta\ll1$ by going beyond the linear response approximation~\cite{PhysRevB.76.144103, PhysRevB.81.224113}. 
Therefore, at WDM parameters with $r_s>2$ and $\theta\sim 1$, it is  interesting to investigate the possibility of an effective ion-ion attraction due electronic nonlinear effects. Specifically, such a mechanism for the ion--ion attraction would manifest in experiments as a maximum in the static structure factor at small wave numbers \cite{zhandos1, Matsuda}.   

A particularly important finding from Ref.~\cite{dornheim2021density} is the pronounced dependence of the different nonlinear density response functions on the system parameters like the density and temperature, which makes the deliberate probing of nonlinear effects a potentially valuable new tool of diagnostics of WDM experiments~\cite{moldabekov2021thermal}. In the present work, we further extend these efforts by investigating the impact of electronic XC-effects on the nonlinear density response of the warm dense electron gas. From a theoretical perspective, we show that the quadratic and cubic density response functions at the second and third harmonic are directly connected to the three- and four-body correlation functions from many-body theory. This explains the observed deficiency of the random phase approximation (RPA) regarding the description of nonlinear effects even at weak coupling and highlights the need for an improved description based on recent results for the electronic local field correction~\cite{dornheim_ML,Dornheim_PRL_2020_ESA,Dornheim_PRB_2021}. On the one hand, our findings demonstrate that the nonlinear density response of WDM constitutes a challenging benchmark for different theories as no universally reliable XC-functionals yet exist in this regime. On the other hand, this allows for the intriguing possibility of experimental measurements of three- and possibly even four-body correlation functions using already existing experimental facilities.


\section{Theory}
We consider a uniform electron gas that is subject to an external static harmonic perturbation~\cite{Dornheim_PRL_2020,bowen2,moroni,moroni2,groth_jcp,dornheim_pre} of wave-vector $\mathbf{q}$ and amplitude $A$, with the corresponding Hamiltonian being given by
\begin{eqnarray}\label{eq:hamiltonian}
\hat H = \hat H_\textnormal{UEG} + 2 A \sum_{l=1}^N \textnormal{cos}\left( \hat{\mathbf{r}}_l\cdot{\mathbf{q}} \right)\ .
\end{eqnarray}
Note that we assume Hartree atomic units throughout this work. The density response of the electrons at a second wave-vector $\mathbf{k}$ is then defined as
\begin{eqnarray}\label{eq:rho}
\braket{\hat\rho_\mathbf{k}}_{q,A} = \frac{1}{V} \left< \sum_{l=1}^N e^{-i\mathbf{k}\cdot\hat{\mathbf{r}}_l} \right>_{q,A} \ , 
\end{eqnarray}
with $\braket{\dots}_{q,A}$ indicating the parameters of the perturbation in Eq.~(\ref{eq:hamiltonian}). It is easy to see that Eq.~(\ref{eq:rho}) only gives non-zero contributions at the integer harmonics of the perturbation wave-vector, i.e., $\mathbf{k}=\eta\mathbf{q}$ with $\eta\in\mathbb{Z}_{\{0\}}$. Naturally, $\eta=1$ corresponds to the original perturbation and, in first order in the perturbation strength $A$, is described by the linear response function $\chi^{(1)}(q)$,
\begin{eqnarray}\label{eq:first_harmonic}
\braket{\hat\rho_\mathbf{q}}_{q,A} = \chi^{(1)}(q)A + \dots \quad .
\end{eqnarray}
For completeness, we mention that the first omitted term in Eq.~(\ref{eq:first_harmonic}) is cubic in $A$ and has been investigated in Refs.~\cite{Dornheim_PRL_2020,dornheim2021density}. The linear response function is typically expressed as 
\begin{eqnarray}\label{eq:LFC}
\chi^{(1)}(q) = \frac{\chi^{(1)}_0(q)}{1-\frac{4\pi}{q^2}\left[ 1-G(q)\right]\chi^{(1)}_0(q)}\ ,
\end{eqnarray}
where $\chi_0^{(1)}(q)$ is the linear response function of the ideal Fermi gas~\cite{quantum_theory}, and $G(q)$ is the static local field correction containing all XC-effects~\cite{kugler1}. In particular, setting $G(q)=0$ in Eq.~(\ref{eq:LFC}) leads to the well-known RPA, i.e., a mean-field description of the density response in LRT.

The second harmonic, $\eta=2$, is in first order described by the quadratic density response function $\chi^{(2)}(q)$,
\begin{eqnarray}\label{eq:second_harmonic}
\braket{\hat\rho_\mathbf{2q}}_{q,A} = \chi^{(2)}(q)A^2 + \dots \quad ,
\end{eqnarray}
which can be expressed in terms of the LFC as~\cite{dornheim2021density}
 \begin{eqnarray}\label{eq:quadratic_LFC}
     \chi^{(2)}_{\rm LFC}( q) &=&  \chi^{(2)}_{0}( q) \left[1-v(q)\left[1-G(q)\right]\chi^{(1)}_{0}(q)\right]^{-2}\nonumber\\
     & & \times \left[1-v(2q)\left[1-G(2q)\right]\chi^{(1)}_{0}( 2q)\right]^{-1}. \label{eq:chi2_LFC}
 \end{eqnarray}
We stress that $\chi^{(2)}_{\rm LFC}\approx\chi^{(2)}(q)$, as some nonlinear screening terms have been neglected. Still, we note that Eq.~(\ref{eq:quadratic_LFC}) has been shown to be in excellent agreement to benchmark PIMC data in Ref.~\cite{dornheim2021density}.

Finally, the third harmonic ($\eta=3$) scales, in first order, cubically in $A$,
\begin{eqnarray}\label{eq:third_harmonic}
\braket{\hat\rho_\mathbf{3q}}_{q,A} = \chi^{(3)}(q)A^3 + \dots \quad ,
\end{eqnarray}
and the approximate expression in terms of the electronic LFC is given by~\cite{dornheim2021density}
 \begin{eqnarray}\nonumber
     \chi^{(3)}_{\rm LFC}( q) &=&  \chi^{(3)}_{0}( q) \left[1-v(q)\left[1-G(q)\right]\chi^{(1)}_{0}(q)\right]^{-3}\\  & & \times \left[1-v(3q)\left[1-G(3q)\right]\chi^{(1)}_{0}( 3q)\right]^{-1}.\label{eq:chi3_LFC}
 \end{eqnarray}

The response functions $\chi^{(1)}$, $\chi^{(2)}$, and $\chi^{(3)}$ are special cases of more general higher order response or correlation functions.
For instance, the connection between correlation function and response function is well known for the two-particle case, for which it reads
\beq\label{eq:connection_linear}
g_2(12)=iL^>(12)+g^<(11)g^<(22)\,,
\eeq
and which is often used throughout linear response theory.
Here, $g_2(12)=\langle \psi^+(\rv_1t_{1^+})\psi(\rv_1t_1)\psi^+(\rv_2t_{2^+})\psi(\rv_2t_2)\rangle$ is the pair correlation function built from creation and annihilation operators $\psi^+$ and $\psi$ and featuring an average (with respect to the unperturbed system, i.e., $A=0$) using the density operator $\langle \ldots \rangle=\mbox{Tr}\{\rho\ldots\}$. This is equal, in the particle-hole channel, to the correlation function of density fluctuations $L^>(12)=\langle  \delta\rho(1)\delta\rho(2)\rangle$, $\delta\rho(1)=\psi^+(1)\psi(1)-\langle\psi^+(1)\psi(1)\rangle$ plus a product of the independent densities $g^<=\langle\psi^+(1)\psi(1)\rangle$.

More familiar is the description of the dynamic structure factor $S\ofqw$ using the fluctuation-dissipation theorem in linear response in frequency-momentum space, namely
\beq
S\ofqw=\frac{i}{2\pi}L^>\ofqw=\frac{1}{n\pi}\frac{1}{e^{-\beta\omega}-1}\mbox{Im}\chi^{(1)}\ofqw\,,
\label{skw}
\eeq
where the spectral representation $iL^>=\mbox{Im}L^R[1+n_B(\omega)]=\mbox{Im}\chi^{(1)}[1+n_B(\omega)]$ with the Bose function $n_B$ and the retarded density fluctuation-correlation function $L^R=\chi^{(1)}$ was used.
The dynamic structure factor as given in Eq. (\ref{skw}) can be measured very well using x-ray Thomson scattering diagnostics~\cite{siegfried_review,kraus_xrts}. Such measurements give information about the pair correlations but also about higher order correlations as they are entering the linear response function via dynamic local field corrections~\cite{dornheim_dynamic,Hamann_PRB_2020,kugler1}.

The three particle correlation function $g_3$ is connected to the quadratic response function $Y$ whose realization at the second harmonic is $Y(\qv,\qv)=\chi^{(2)}(q)$ and can be used to directly study higher order exchange and correlation
instead of indirectly trying to gauge their influence by measuring or calculating the difference to the mean field RPA values in linear response. In detail, this connection is given by
\bea\label{eq:connection_quadratic}
\lefteqn{g_3(123)=-Y(123)}&&\nonumber\\
&&-iL(12)g(33)-iL(13)g(22)-iL(23)g(11)\nonumber\\
&&+g(11)g(22)g(33)\,.
\eea
Specifically, pure three-body correlations are accounted for by the correlation function of three density fluctuations $Y=\langle\delta\rho(1)\delta\rho(2)\delta\rho(3)\rangle$, which is readily identified as the quadratic response function. Additional contributions come from pair-correlations multiplied with single particle states and the term with three independent states.

The quadratic fluctuation dissipation theorem connects the quadratic response function with the dynamic quadratic structure factor \cite{Kalman_1987,PhysRevE.54.3518}
\begin{align}
&S(\qv_1,\omega_1;\qv_2,\omega_2)=\frac{-4}{n}\frac{1}{D}\Big\{
\left(1-e^{-\beta\omega_1} \right)e^{-\beta\omega_2}
\\
&
 \qquad\qquad\quad\times\left[\mathscr{Y}'(\qv_1,\omega_1;\qv_2,\omega_2)-\mathscr{Y}'(\qv_2,\omega_2;-\qv,-\omega)\right]\nonumber\\
&+\left(1-e^{-\beta\omega_2}\right)
\left[\mathscr{Y}'(\qv_1,\omega_1;\qv_2,\omega_2)-\mathscr{Y}'(-\qv,-\omega;\qv_1,\omega_1)\right]\! \Big\}\;,
\nonumber
\end{align}
with the argument $\qv=\qv_1+\qv_2$ and similarly for $\omega$. In addition, $\mathscr{Y}'$ denotes the real part of the retarded quadratic response function $Y^R$.

For completeness, we also give the connection between the quadruple correlation function and the cubic response function,
\bea
\lefteqn{g_4(1234)=iZ(1234)}&&\nonumber\\
&&-Y(123)g(44)-Y(124)g(33)\nonumber\\
&&-Y(134)g(22)-Y(234)g(11)\nonumber\\
&&-L(12)L(34)-L(13)L(24)-L(14)L(23)\nonumber\\
&&-iL(12)g(33)g(44)-iL(13)g(22)g(44)\nonumber\\
&&-iL(14)g(22)g(33)-iL(23)g(11)g(44)\nonumber\\
&&-iL(24)g(11)g(33)-iL(34)g(11)g(2')\nonumber\\
&&+g(11)g(22)g(33)g(44)\;.
\eea
One observes, as with the triple correlation before, that the total four-particle correlation function is given by direct four-particle correlations [the cubic response function $Z$, which at the third harmonic is $Z(\qv,\qv,\qv)=\chi^{(3)}(q)$], plus a sum of pair- and triple-particle correlations with the remaining particle(s) being factored out, and of course the ideal case of four independent particles. 

\section{Results}

\begin{figure*}\centering
\includegraphics[width=0.4285\textwidth]{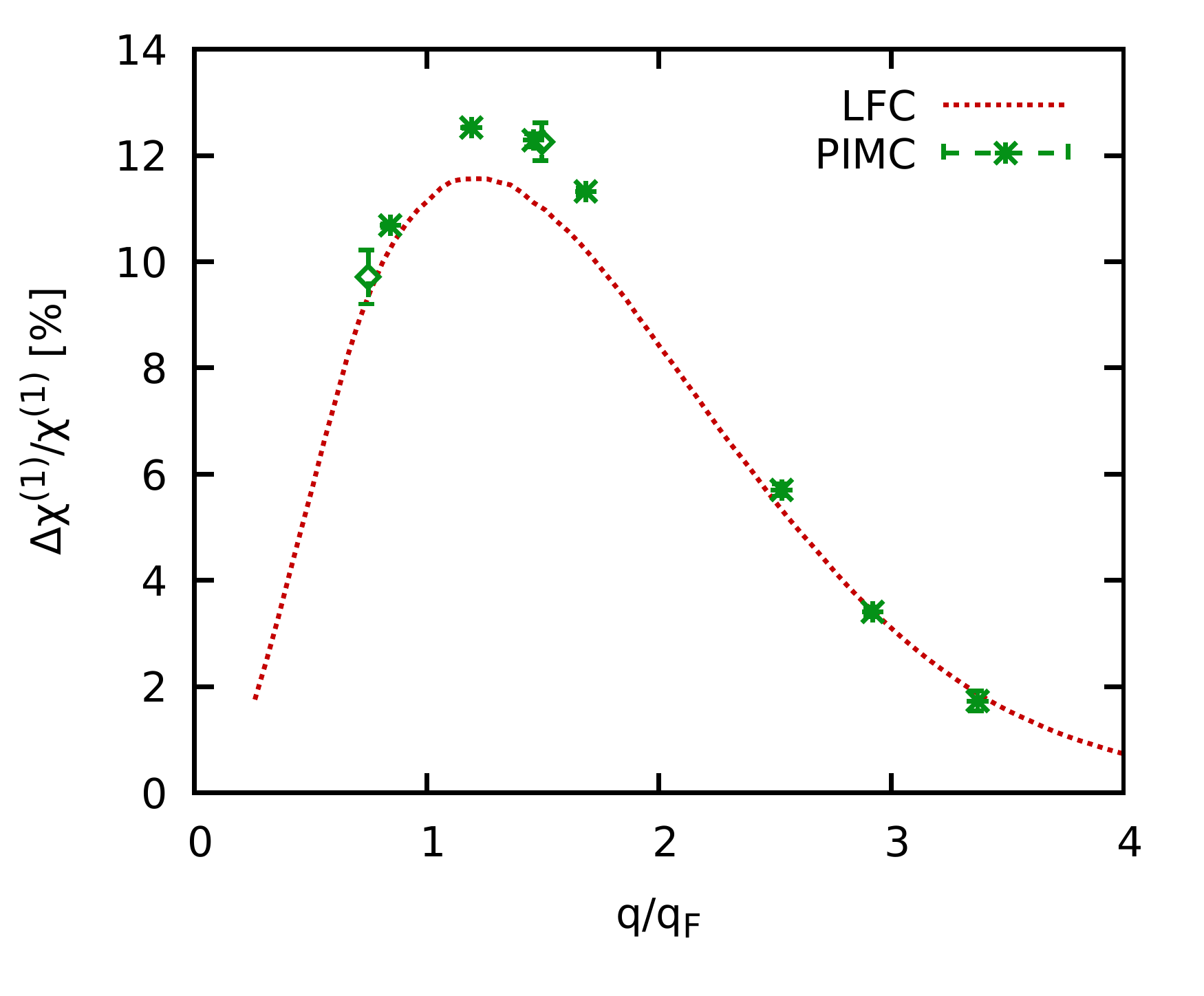}
\includegraphics[width=0.4285\textwidth]{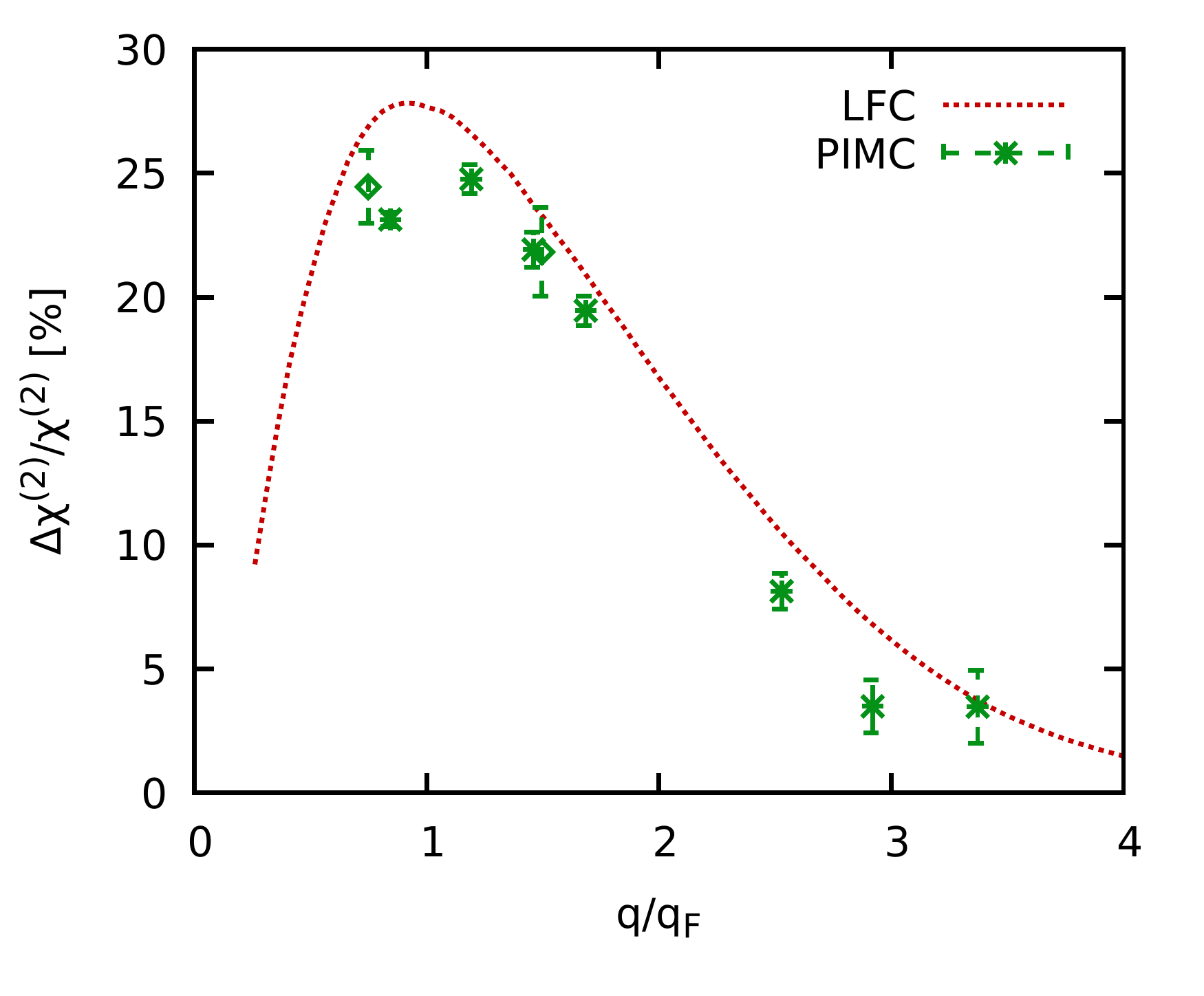}
\caption{\label{fig:PIMC}Relative difference between a correlated response function $\chi$ and the mean-field analogue $\chi_\textnormal{RPA}$ for the unpolarized UEG at $r_s=2$ and $\theta=1$. The left and right panels show results for the linear response function $\chi^{(1)}(q)$ and the quadratic response function $\chi^{(2)}(q)$. The PIMC data (green symbols) have been taken from Ref.~\cite{dornheim2021density}, and the dotted red curves have been computed using as input the LFC from Ref.~\cite{dornheim_ML}.
}
\end{figure*}

The central property that we consider in this work is the relative difference between the density response functions $\chi^{(\eta)}(q)$ defined in Eqs.~(\ref{eq:first_harmonic})-(\ref{eq:third_harmonic}) above evaluated in RPA and using a LFC,
\begin{eqnarray}\label{eq:delta}
   \frac{ \Delta \chi^{(\eta)}(q) }{ \chi^{(\eta)}(q)} = \frac{\chi_{\rm LFC}^{(\eta)}(q)-\chi^{(\eta)}_{\rm RPA}(q)}{\chi_{\rm LFC}^{(\eta)}(q)}\ .
\end{eqnarray}
More specifically, we use the recent neural-net representation of $G(q;r_s,\theta)$ by Dornheim \textit{et al.}~\cite{dornheim_ML,dornheim2021machinelearning} that is based on extensive PIMC data and is accurate over the entire relevant parameter range.
The results of Eq.~(\ref{eq:delta}) are shown as the dotted red lines in Fig.~\ref{fig:PIMC} for $\eta=1$ (left panel) and $\eta=2$ (right panel) for $r_s=2$ and $\theta=1$. For completeness, we mention that such conditions are relevant for contemporary WDM research and can be realized experimentally for example with aluminum~\cite{Sperling_PRL_2015,Ramakrishna_PRB_2021}. First and foremost, we note that the impact of XC-effects vanishes both in the limit of large- and small wave numbers and attains a maximum around $q=q_\textnormal{F}$ for both harmonics. In particular, it is well known that XC effects do not play a role for $q\to0$ due to the perfect screening in an electron gas~\cite{kugler_bounds}. In the limit of large wave numbers, on the other hand, the density response of the electrons is fully governed by single-particle effects, which, too, means that XC-effects become unimportant.
Let us next consider the green data points that have been obtained using the PIMC method; see Ref.~\cite{dornheim2021density} for details. In particular, the stars and diamonds have been obtained for $N=14$ and $N=20$ electrons, but cannot be distinguished within the given Monte Carlo error bars. In addition, the data points are in good agreement to the LFC-curves, which illustrates the high quality of the expressions in Eqs.~(\ref{eq:LFC})-(\ref{eq:chi3_LFC}).
Finally, we observe that the impact of XC-effects is larger by more than a factor of two for the quadratic response shown in the panel on the right. This is a direct consequence of the respective relations of $\chi^{(1)}$ and $\chi^{(2)}$ to the two- and three-body correlation functions [cf.~Eqs.~(\ref{eq:connection_linear}) and (\ref{eq:connection_quadratic}) above], and constitutes the central result of this work.

\begin{figure*}\centering
\includegraphics[width=0.4285\textwidth]{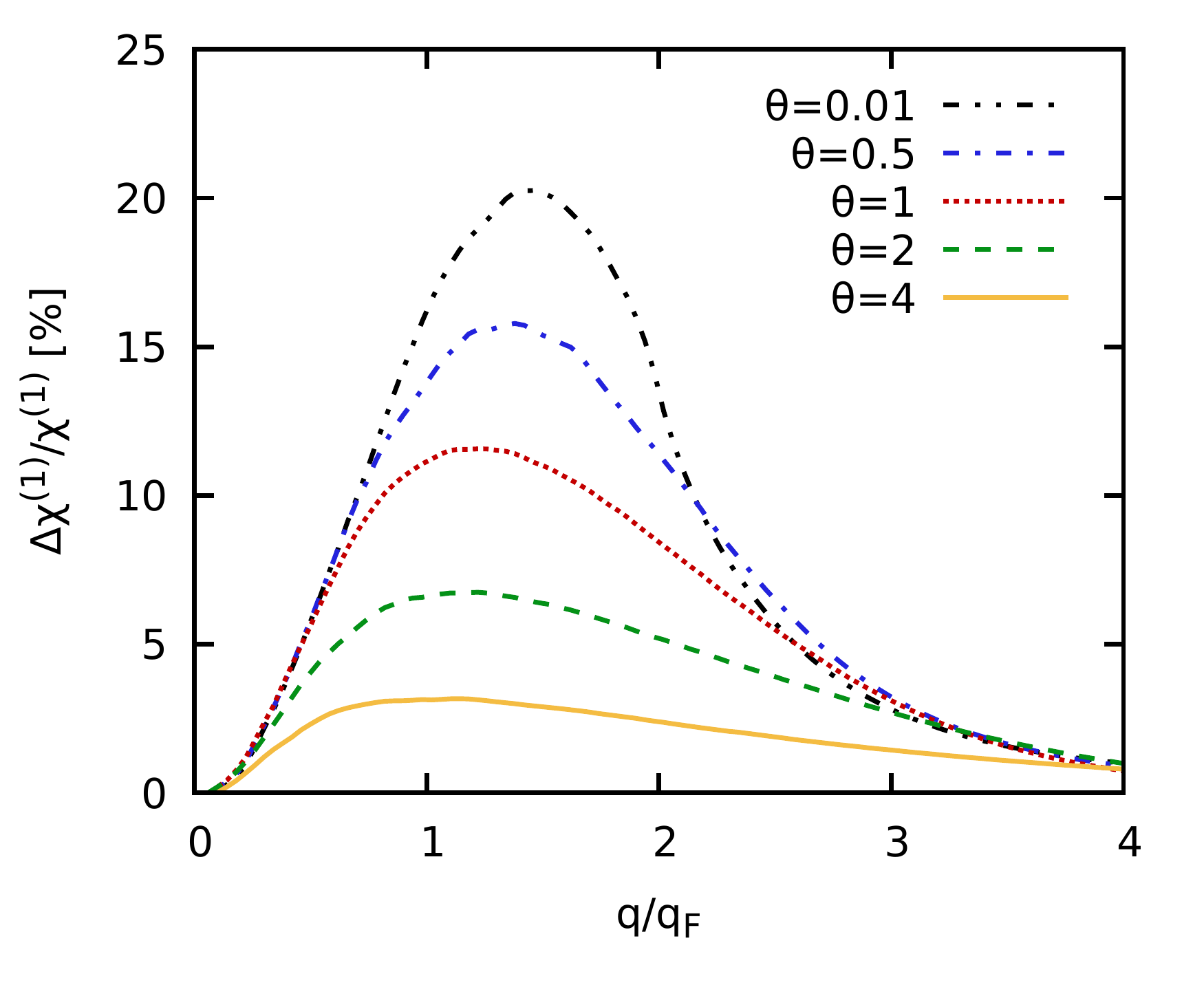}
\includegraphics[width=0.4285\textwidth]{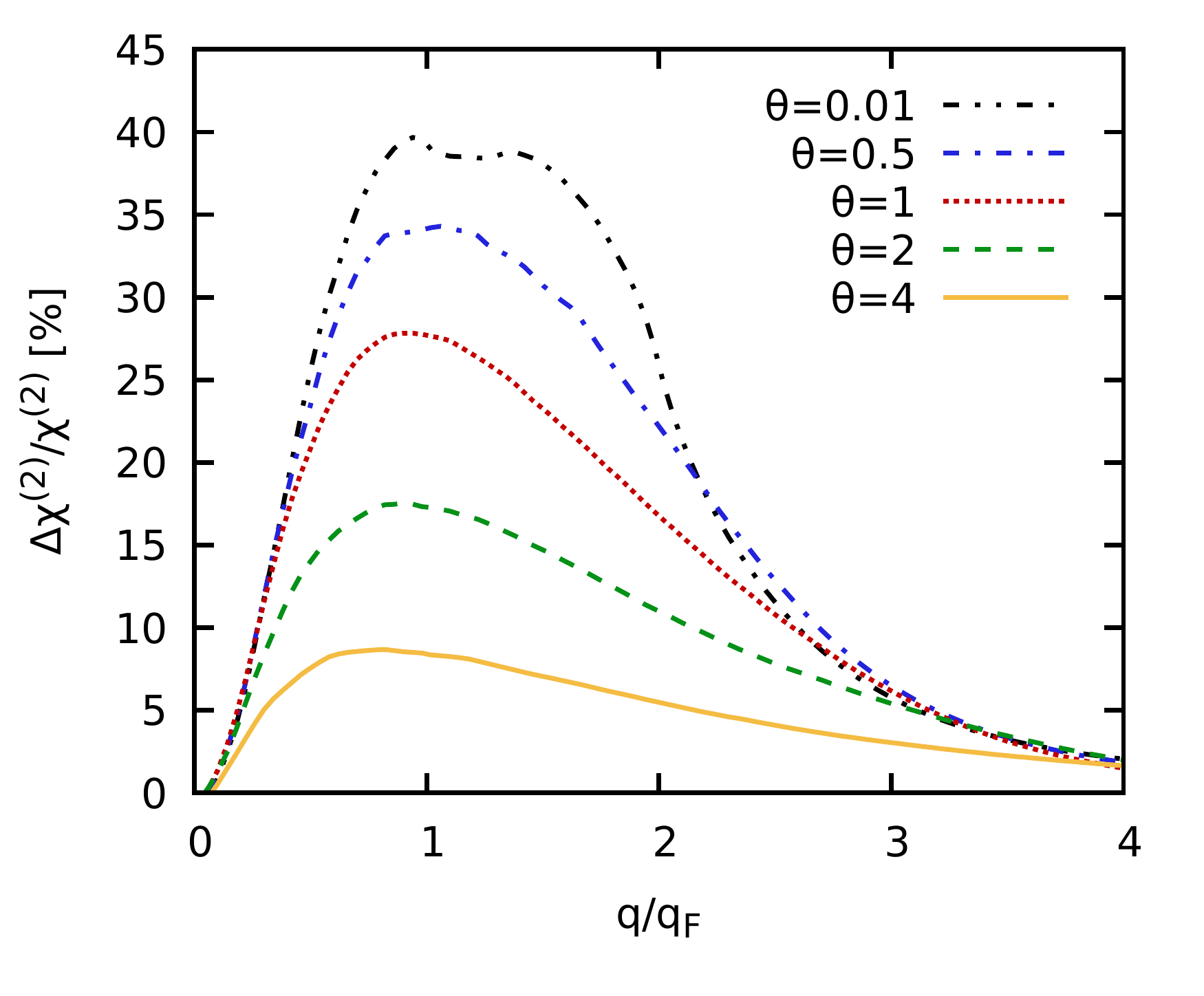}\\\vspace*{-0.8cm}
\includegraphics[width=0.4285\textwidth]{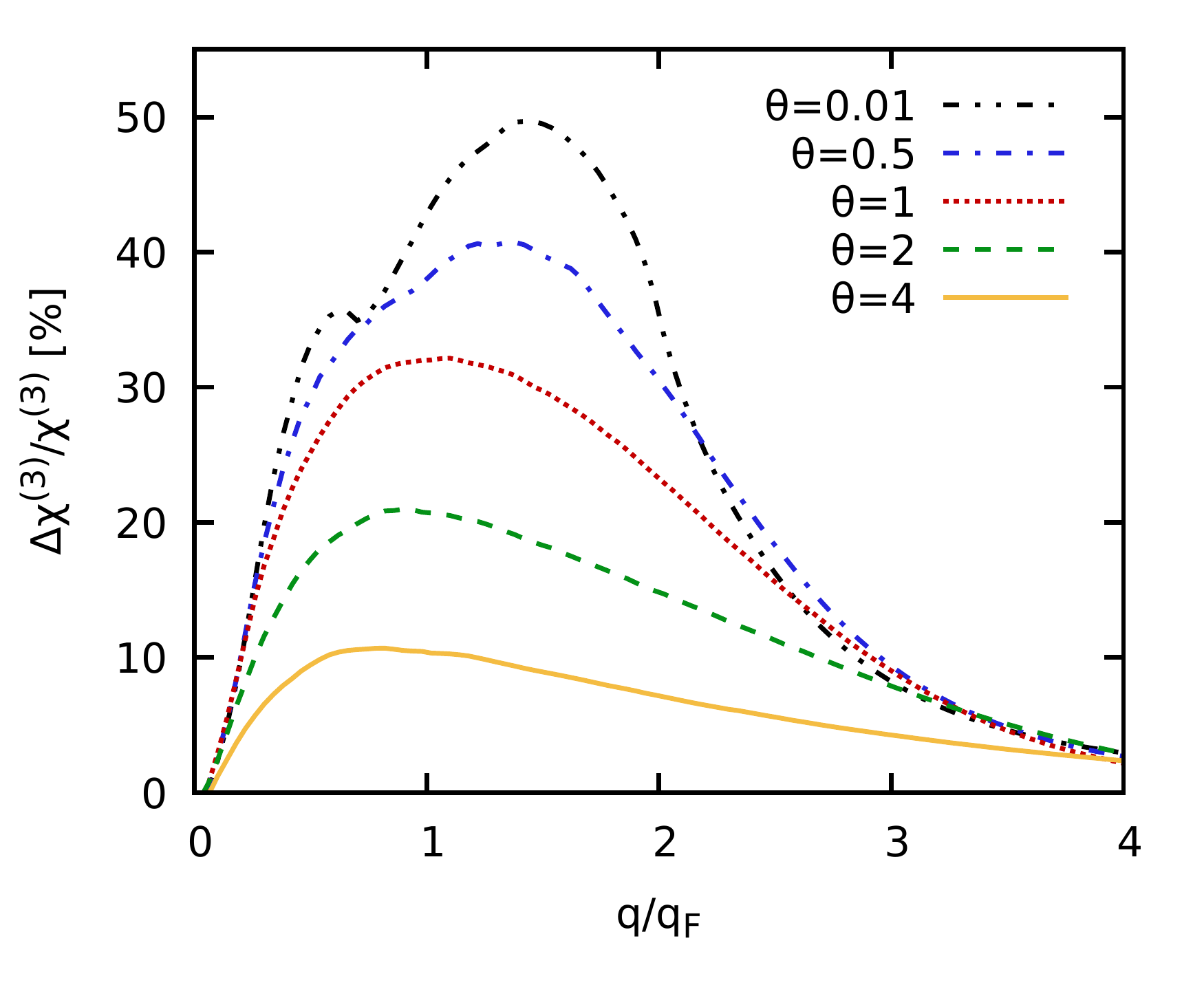}\includegraphics[width=0.4285\textwidth]{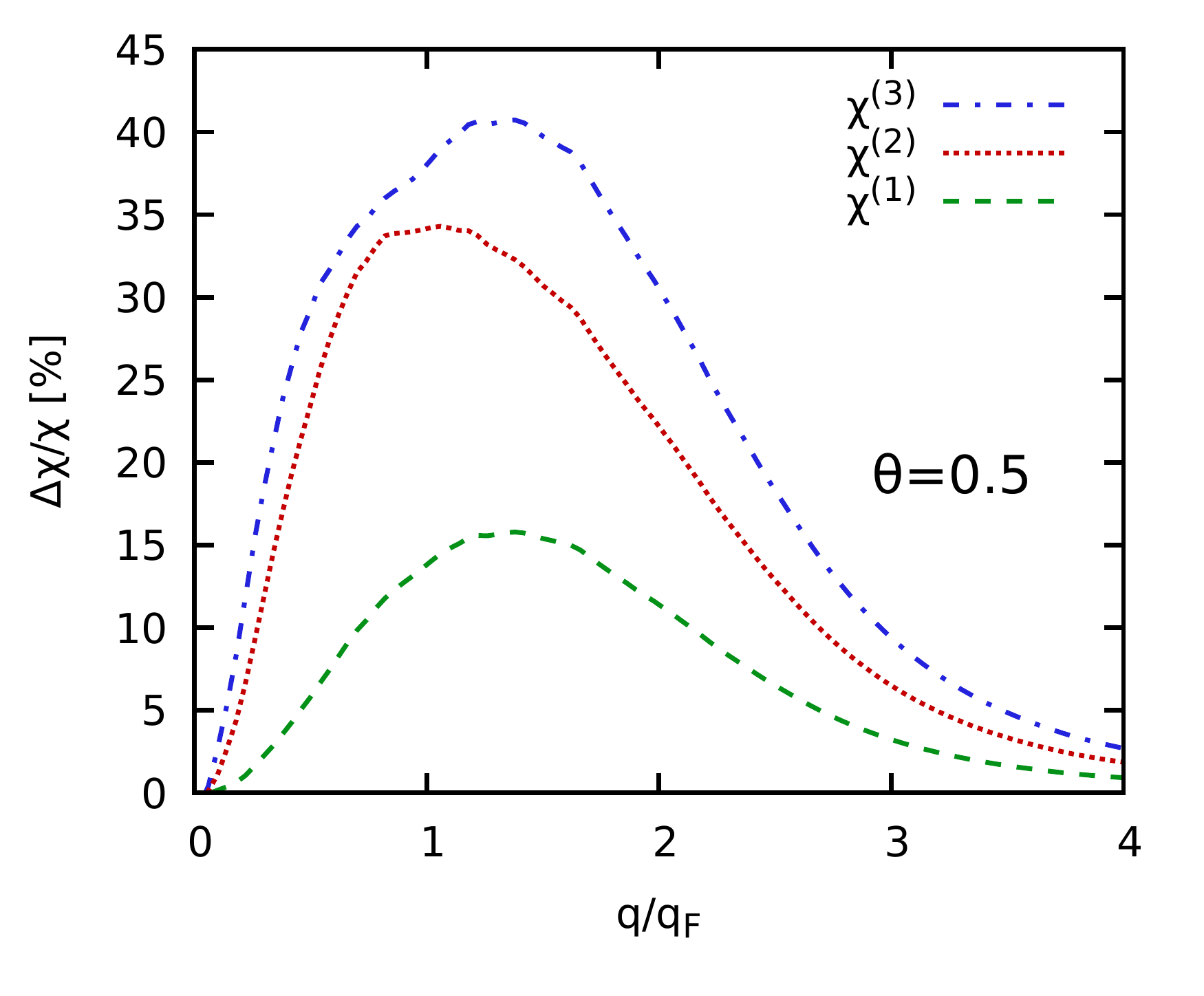}
\caption{\label{fig:theta}Relative difference between a correlated response function $\chi$ (using as input the LFC from Ref.~\cite{dornheim_ML}) and the mean-field analogue $\chi_\textnormal{RPA}$ [cf.~Eq.~(\ref{eq:delta})] for the unpolarized UEG at $r_s=2$.
}
\end{figure*} 

A more systematic investigation of this effect is shown in Fig.~\ref{fig:theta}, where we evaluate Eq.~(\ref{eq:delta}) for $\eta=1$ (top left), $\eta=2$ (top right), and $\eta=3$ (bottom left) at $r_s=2$ for five different temperatures. A comparison between $\eta=1-3$ gives a qualitatively similar behaviour in all three cases: for the lowest temperature ($\theta=0.01$, dash-double-dotted black line), the system is most strongly correlated and the impact of XC-effects is most pronounced; with increasing $\theta$, the curves become less dependent on the wave number, and the peaks are decreased in magnitude. In addition, we find that $\Delta\chi^{(\eta)}/\chi^{(\eta)}(q)$ systematically increases with $\eta$ over the entire depicted temperature-range. For example, the impact of XC effects on the LRT function in the ground state is approximately $20\%$, but $\sim40\%$ for the quadratic response at the second harmonic, and exceeds $50\%$ for the cubic response at the third harmonic.
The same trend is more explicitly shown in the bottom right panel, where we directly compare Eq.~(\ref{eq:delta}) for different $\eta$ with $\theta=0.5$ being fixed. Again, the impact of XC-effects is least pronounced for the linear response function (dashed green), with a maximum impact of $\sim15\%$. The quadratic response function (dotted red) is changed by the LFC by almost $35\%$, and exceeds the green curve for all values of the wave number $q$. Lastly, the cubic response function (dash-dotted blue) is most strongly affected by XC-effects, with a maximum exceeding $40\%$. Again, the order of the response function holds for all $q$.

This strongly reaffirms our earlier point about the importance of correlation effects in the description of the nonlinear density response of WDM due to their relation to three- and four-body correlation functions.

\begin{figure*}\centering
\includegraphics[width=0.4285\textwidth]{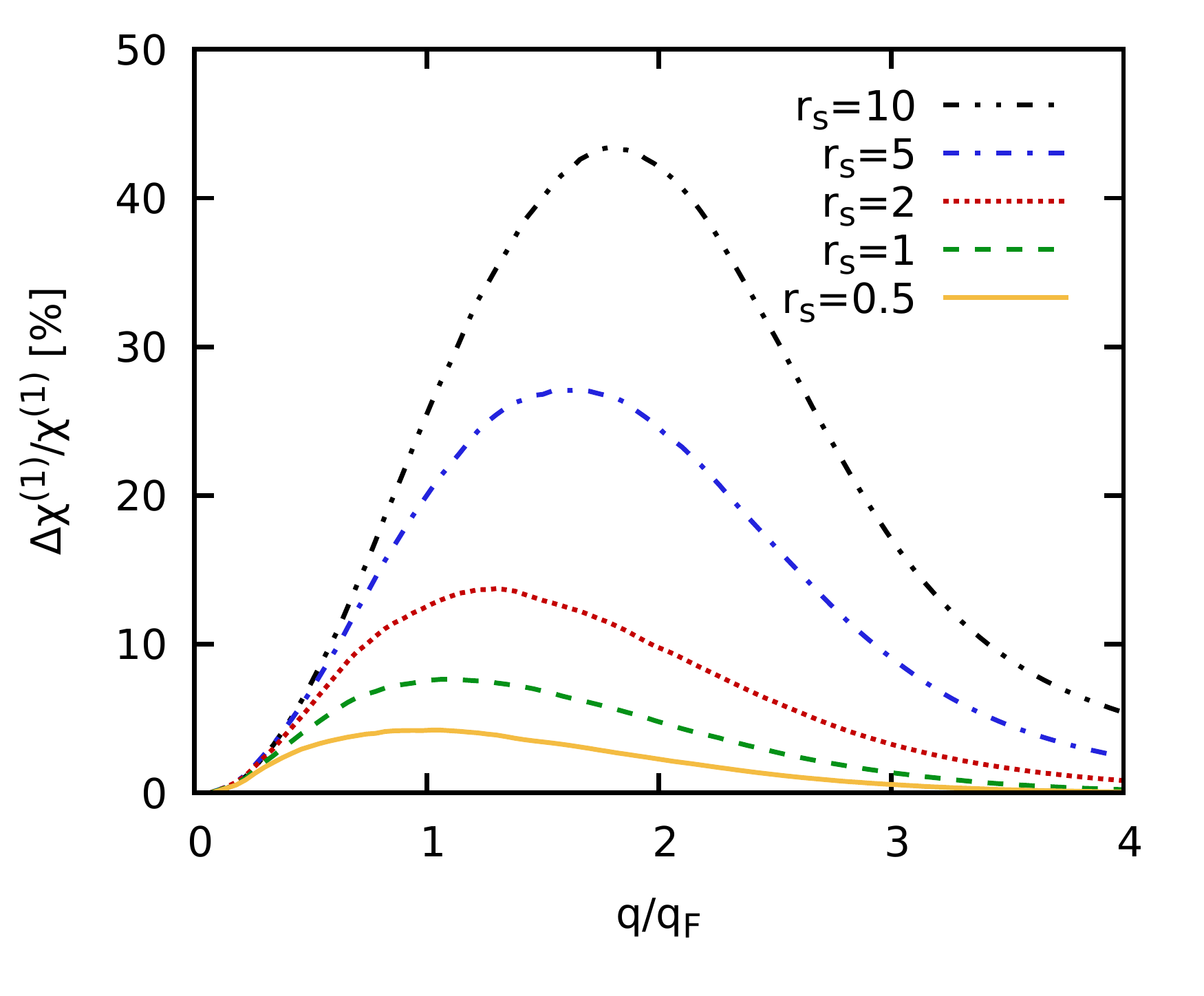}
\includegraphics[width=0.4285\textwidth]{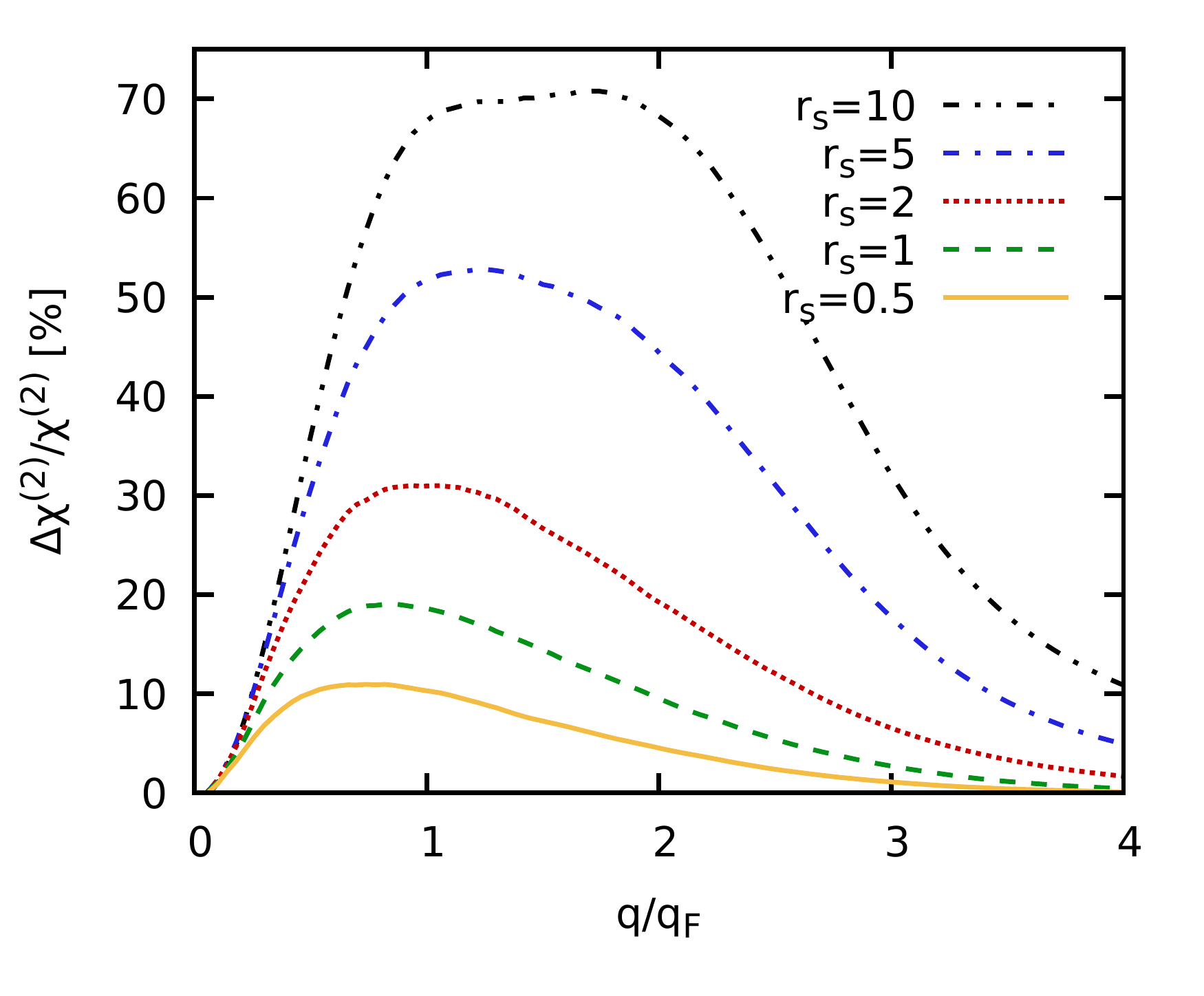}\\\vspace*{-0.8cm}
\includegraphics[width=0.4285\textwidth]{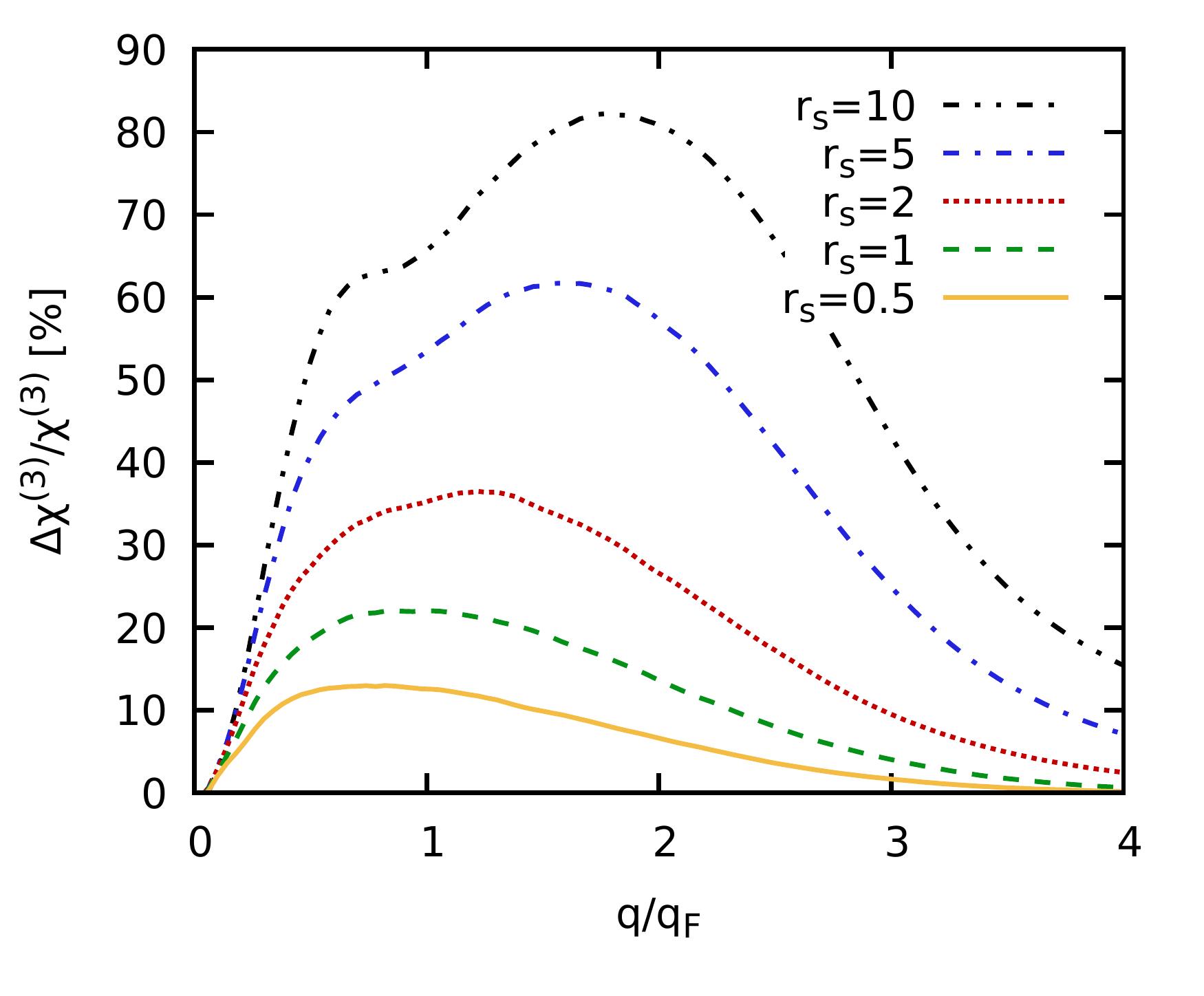}\includegraphics[width=0.4285\textwidth]{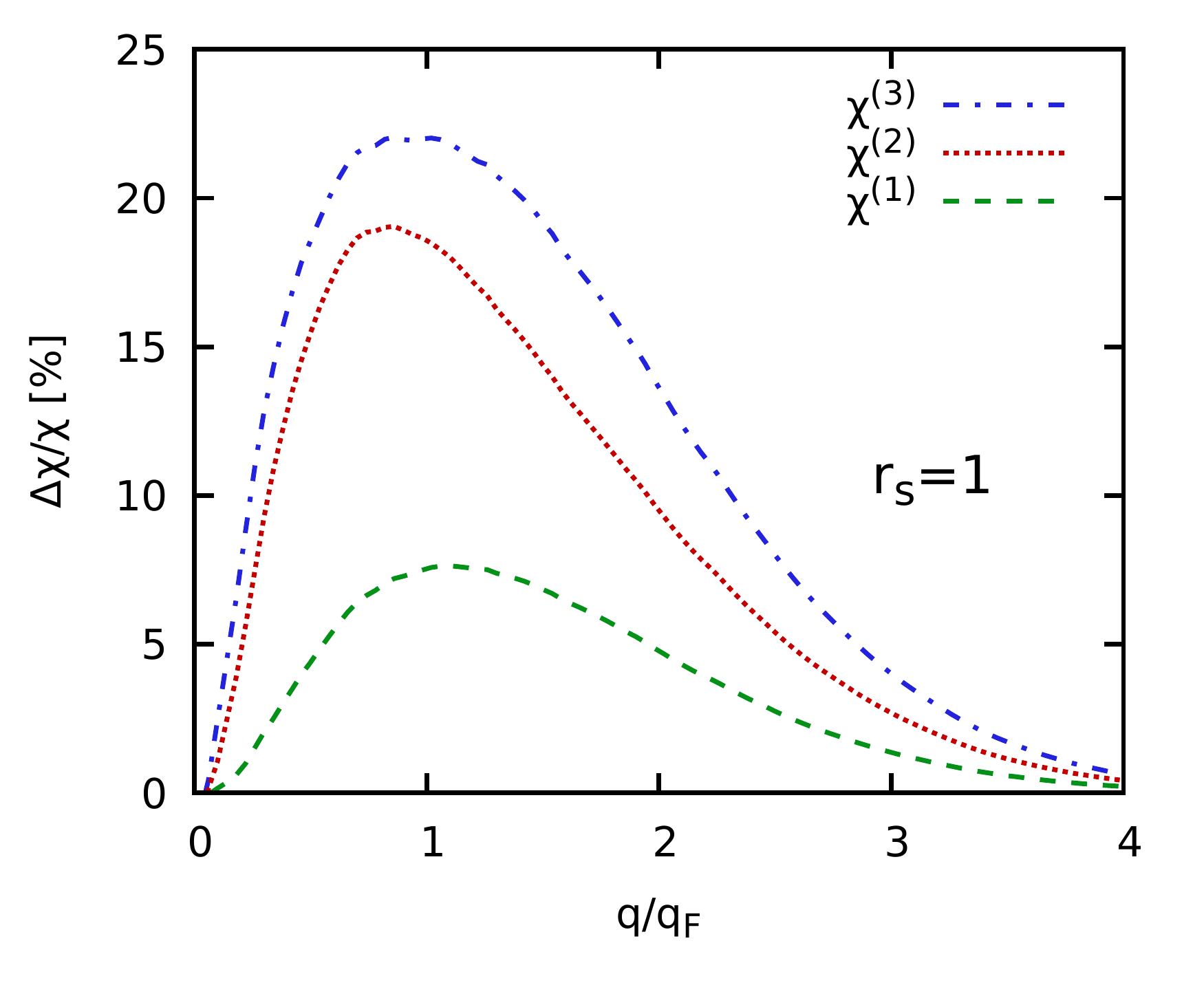}
\caption{\label{fig:rs}Relative difference between a correlated response function $\chi$ (using as input the LFC from Ref.~\cite{dornheim_ML}) and the mean-field analogue $\chi_\textnormal{RPA}$ [cf.~Eq.~(\ref{eq:delta})] for the unpolarized UEG at $\theta=0.75$.
}
\end{figure*}

A further interesting topic of investigation is the dependence of the impact of XC-effects on the density. To this end, we evaluate Eq.~(\ref{eq:delta}) at $\theta=0.75$ for five different values of $r_s$. At $r_s=0.5$, the density is high and, consequently, the system is only weakly coupled~\cite{quantum_theory,Ott2018}. Indeed, we find that the LFC impacts the linear response function by less than $5\%$ at the Fermi wave number, where the impact is most pronounced. In contrast, the LFC changes both $\chi^{(2)}(q_\textnormal{F})$ and $\chi^{(3)}(q_\textnormal{F})$ by more than $10\%$. This vividly illustrates the challenging nature of an accurate theory of the nonlinear density response even at weak coupling, where the RPA is relatively accurate for the linear response~\cite{review,dornheim_HEDP}. Upon increasing the density parameter, the system becomes more strongly coupled and, consequently, the change due to the LFC monotonically increases. At $r_s=10$, which is located at the margin to the electron liquid regime~\cite{dornheim_dynamic,dornheim_electron_liquid}, the RPA breaks down for all $\eta$, with a maximum error of $\gtrsim40\%$ for the linear response, $\gtrsim70\%$ for the quadratic response, and $\gtrsim80\%$ for the cubic response.

A more direct comparison is shown in the bottom right panel of Fig.~\ref{fig:rs} for $r_s=1$ being constant. Again, the ordering of the curves is monotonous with $\eta$ over the entire wave-number range, as it is by now expected.

\section{Discussion}

In this work, we have examined the relation between the nonlinear density response and higher order correlation functions in WDM. More specifically, the linear response of a system is connected to the pair correlation function, the quadratic response to three-body correlations, and the cubic response to four-body correlations. This, in turn, directly implies the increasing importance of electronic XC-effects with the order of the external perturbation.

This prediction has been verified by investigating the relative impact of the LFC on the linear response function $\chi^{(1)}(q)$ at the first harmonic, the quadratic response function $\chi^{(2)}(q)$ at the second harmonic, and the cubic response function $\chi^{(3)}(q)$ at the third harmonic for different densities and temperatures. Indeed, our results unambiguously indicate that XC-effects become increasingly important with the order of the nonlinear response and, thus, with the order of the many-body correlation function to which they are related. Remarkably, we find this trend to hold for all densities and temperatures, and over the entire range of relevant wave numbers.

In other words, the accurate description of the nonlinear density response is nontrivial even at high density and weak coupling, where RPA is only applicable to the linear response of the system. This makes it a suitable and challenging benchmark for the development of new theoretical approaches, such as the introduction of new XC-functionals for DFT or improved self-energies for nonequilibrium Greens functions~\cite{Schl_nzen_2019}.

On the other hand, our work directly points to the unprecedented possibility of a straightforward experimental observation of higher-order many-body effects by deliberately probing a WDM sample in the nonlinear regime. For example, sufficient perturbation strengths can already be realized with free electron lasers using the novel seeding technique~\cite{Fletcher2015,Amann2012}.

\acknowledgments
\section{Acknowledgements}
We gratefully acknowledge fruitful discussions with Michael Bonitz.

This work was partly funded by the Center for Advanced Systems Understanding (CASUS) which is financed by Germany's Federal Ministry of Education and Research (BMBF) and by the Saxon Ministry for Science, Culture and Tourism (SMWK) with tax funds on the basis of the budget approved by the Saxon State Parliament.
The PIMC calculations were carried out at the Norddeutscher Verbund f\"ur Hoch- und H\"ochstleistungsrechnen (HLRN) under grant shp00026, and on a Bull Cluster at the Center for Information Services and High Performance Computing (ZIH) at Technische Universit\"at Dresden.

\bibliographystyle{iopart-num}
\bibliography{bibliography2.bib}

\end{document}